\documentclass[]{aa}
\usepackage{graphicx}
\include{psfig}

\begin{document}

\title{{\sl FUSE} search for 10$^5$--10$^6$~K gas in the rich clusters of 
galaxies Abell~2029 and Abell~3112}


\authorrunning{A. Lecavelier et al.}
\titlerunning{Search for 10$^5$--10$^6$~K gas in Abell~2029 and Abell~3112} 

\author{
A.~Lecavelier des Etangs \inst{1} 
\and
Gopal-Krishna \inst{2} 
\thanks{ Alexander von Humboldt Fellow (On leave from NCRA.TIFR,
Pune, India) }
\and
F.~Durret \inst{1} \\
}

\offprints{A. Lecavelier des Etangs,
            \email{lecaveli@iap.fr}}

\institute{Institut d'Astrophysique de Paris, CNRS, 98 bis bld~Arago, 
F-75014 Paris, France
         \and
Max-Planck-Institut f\"ur Radioastronomie, Auf dem H\"ugel 69, 
53121 Bonn, Germany
}

\date{Received ...; accepted ...} 

\abstract{Recent Chandra and XMM X-ray observations of rich clusters
of galaxies have shown that the amount of hot gas which is cooling
below $\sim$~1 keV is generally more modest than previous
estimates. Yet, the real level of the {\it cooling flows}, if any,
remains to be clarified by making observations sensitive to different
temperature ranges. As a follow-up of the {\sl FUSE} observations
reporting a positive detection of the O{\sc vi} doublet at 1032,
1038\AA\ in the cluster of galaxies Abell~2597, which provided the
first direct evidence for $\sim 3\ 10^5$~K gas in a cluster of
galaxies, we have carried out sensitive spectroscopy of two rich
clusters, Abell~2029 and Abell~3112 ($z \sim 0.07$) located behind low
HI columns. In neither of these clusters could we detect the O{\sc vi}
doublet, yielding fairly stringent limits of
$\sim$27~M$_\odot$yr$^{-1}$ (Abell~2029) and
$\sim$25~M$_\odot$yr$^{-1}$ (Abell~3112) to the cooling flow rates
using the 10$^5$--10$^6$~K gas as a tracer. The non-detections support
the emerging picture that the cooling-flow rates are much more modest
than deduced from earlier X-ray observations.

}

\maketitle
%

\section{Introduction}

The concept of cooling flows, arising from simple physical
considerations applied to the X-ray data, has provided an important
input to the models of galaxy formation ({\it e.g.,} Fabian 1994; 
Mathews \& Brighenti 2003).  On the other hand, the expected outcome, often
amounting to $\geq 10^{10}$\,M$_\odot$ of cooled gas inside the cores of
many rich clusters, has been rather elusive, despite sensitive
searches made in multiple wavebands ({\it e.g.,} Donahue \& Voit, 2003 and
references therein). Evidence does exist for some such cooler gas in
the form of highly extended optical nebulosities seen in the cores of
several cooling-flow clusters ({\it e.g.,} Hu et al. 1985; Heckman 1985), or CO
emission (Edge \& Frayer 2003; Salom\'e \& Combes 2003).  However,
the masses involved are tiny in comparison with the simple predictions
of the cooling-flow rates.  The most recent addition to this intrigue
comes from the high resolution spectroscopic observations of several
cooling flows with XMM-Newton and Chandra which have revealed a clear
deficit of gas at $\leq$1 keV in comparison with the prediction of
simple cooling-flow models ({\it e.g.,} Peterson et al. 2001, 2003; Kaastra
et al. 2001; Tamura et al. 2001).  In this context, spectroscopic observations 
to search for the UV resonance lines of O\,{\sc vi} 1032,1038\,\AA\ 
emission doublet, a reliable tracer of gas in the temperature range $10^5 -
10^6$K, can play an important role. In fact, using the Far Ultraviolet Spectroscopic
Explorer ({\sl FUSE}), Oegerle et al. (2001)
have reported a convincing detection of O\,{\sc vi} $\lambda$1032\AA\ line 
emission from the core of the rich cluster Abell~2597, and have thus estimated
a rate of $\sim$40\,M$_\odot$yr$^{-1}$ of intra-cluster medium (ICM) in this cluster cooling 
through $\sim$10$^6$\,K. This rate is only a third of the value inferred from 
the analysis of ROSAT data (Sarazin et al. 1995). For the other strong
cooling-flow candidate, Abell~1795, the 
({\sl FUSE}) observations by Oegerle et al. (2001) 
placed an upper limit of 28\,M$_\odot$yr$^{-1}$ within the
central 28~kpc region. It may be noted that the cD galaxies in the
cores of both these clusters host twin-jet radio sources extending on a 
100~kpc scale and having radio luminosities typical of cluster radio
sources in the nearby universe. Also, no O\,{\sc vi} has been
detected by {\sl FUSE} in the Virgo and Coma clusters (Dixon et al. 2001a).

Motivated by the above, partly successful attempt using {\sl FUSE}, we
have carried out fairly sensitive {\sl FUSE} spectroscopy of another
two nearby (z$\sim$0.07) BM type~I clusters of richness class~2,
namely, Abell~2029 and Abell~3112. For both clusters, the then
available estimates of the cooling-flow rates were mostly in excess of
300\,M$_\odot$yr$^{-1}$, based on ROSAT and/or ASCA data (Peres et
al. 1998; Sarazin et al. 1998).  A crucial advantage with these two
clusters is the exceptionally low H\,{\sc i} foreground columns
(Table~1), which is an essential pre-requisite for undertaking a
sensitive search in the far ultraviolet. In the meanwhile, based on
progressively new X-ray observations, the derived properties of these
clusters (like many others), such as cooling-flow rates and the radial
profile of the ICM temperature, have undergone a sharp ``evolution''
(Sect.~2), making it desirable to employ independent observational
probes for addressing this issue.  We report here one such attempt,
based on {\sl FUSE} observations of the two clusters.

\section{The two clusters Abell~2029 and Abell~3112}

\subsection{Abell~2029}

Abell~2029, at a redshift $z$=0.0767, is one of the most optically
regular rich clusters known, dominated by a central cD galaxy
(Dressler 1978). It is one of the brightest X-ray clusters (David et
al. 1993), with a luminosity $L_X$(2-10\,keV)=1.1$\times
10^{45}h_{70}^{-2}$~erg/s (Lewis et al. 2002), an X-ray gas
temperature of 9.4~keV (Castillo-Morales \& Schindler 2003) and a
regular structure in X-rays as well. Recently, Chandra observations
have revealed a highly regular and smooth X-ray morphology without any
excess emission near the center, as well as a radial increase in the
temperature from T$\sim$3~keV within the inner $\sim$10~kpc to
T$\sim$9~keV near $r$=250~kpc (Lewis et al. 2002). It is noteworthy
that an opposite radial dependence had been found in earlier X-ray
studies of this cluster, based on ASCA/ROSAT data (Sarazin et
al. 1998; Irwin et al. 1999). In another study, based on an improved
analysis of ASCA observations, White (2000) found consistency with no
radial variation of temperature in this cluster.  Chandra observations
have provided no evidence for a cooling flow in this cluster and led
to an estimated mass deposition rate of $0.0\pm 0.7$\,M$_\odot$yr$^{-1}$
(Lewis et al. 2002), consistent with the trend revealed by XMM-Newton
spectroscopy (Kaastra et al. 2001; Boehringer et al. 2002). 
Note that even though the Chandra measurement only refers to the
central 5$\arcsec$\ region, the value obtained after integrating over 
the cooling radius would still be very modest and well within the 
1$\sigma$ upper limit of $\dot{M}$=175~M$_\odot$\,yr$^{-1}$
deduced by White (2000).
On the other hand, such estimates are in marked contrast to the large mass
deposition rate of $\dot{M}\simeq 400-600$\,M$_\odot$\,yr$^{-1}$ (Edge
et al. 1992; Peres et al. 1998), and to the somewhat lower value of
$162^{+30}_{-26}$~M$_\odot$yr$^{-1}$ estimated by Allen \& Fabian
(1997) using ROSAT data.
However, it is in agreement with the upper
limit on molecular hydrogen derived from CO observations by Salom\'e
\& Combes (2003). The radio source associated with the ultra-luminous
central cD galaxy of Abell~2029 has a very steep radio spectrum
($\alpha$=$-1.5$ above $\sim$100~MHz) and has been discussed in detail
by Taylor et al. (1994).  The two radio lobes exhibit a pronounced
``wide-angle-tail (WAT)'' morphology with an overall extent of at
least 80~kpc (H$_o$=50~km~s$^{-1}$~Mpc$^{-1}$ used throughout this
paper). The compact central core has an inverted radio spectrum
($\alpha$=+0.2); the spectrum steepens to $\alpha$=$-2.7$ near the
edges of the radio tails. From synchrotron ageing arguments, these
authors estimated an age of $1.2\times 10^7$~years. From the extremely
large rotation measure values reached within $\sim$10~kpc of the radio
nucleus ($|$RM$|$ up to 8000~rad~m$^{-2}$) and from its observed
distribution, Taylor et al. have also inferred an ICM magnetic field
of $> 0.2 \mu$G, ordered on a scale of 20-100~kpc.  Considering that
the observed ultra-steep radio spectrum and very large rotation
measure are both strongly suggestive of a cooling-flow environment
(e.g., Taylor et al. 1994), the non-detection of [O\,{\sc ii}] 
emission and the absence of blue colors in the cluster core are
surprising (cf. McNamara \& O'Connell 1989). It is likewise intriguing
that the source exhibits a clear WAT morphology, which is usually
believed to be associated with the ICM winds resulting from an ongoing
cluster merger ({\it e.g.,} Loken et al. 1995; Roettiger et al. 1996),
which is clearly at odds with the smooth and regular X-ray surface
brightness distribution of this cluster (see Lewis et al. 2002).

\subsection{Abell~3112}

Abell~3112 is an optically regular cluster at $z$=0.0746 showing no
evidence for substructure (Biviano et al. 2002). It appears also quite
smooth and dynamically relaxed in X-rays, with some distortions near
the center. This Chandra image also shows an X-ray point source
coincident with the cD and with the core of a powerful radio source
(Takizawa et al. 2003).

In contrast to the image analyses of EXOSAT and ROSAT data which
yielded mass deposition rates of $\sim 400$~M$_\odot$\,yr$^{-1}$ 
(Edge et al. 1992; Allen \& Fabian 1997; Peres et al. 1998), a much
smaller value of $44^{+52}_{-32}$~M$_\odot$yr$^{-1}$ has been derived from
Chandra data by Takizawa et al. (2003). The distortions seen near the
center in the Chandra image probably signify
a dynamical interaction between the ICM and the powerful radio source
hosted by the central cD galaxy. The nuclear X-ray point source with a
radio counterpart is characterized as a strongly absorbed AGN
(Takizawa et al. 2003).  Using these data, Takizawa et al. have
further deduced a radially increasing temperature from $\sim$2~keV at
the center to $\sim$6~keV near $r$=150~kpc, in broad agreement with
the ASCA result (White 2000). However, as in the case of Abell~2029,
such a temperature gradient is in conflict with the trend inferred
from ROSAT/ASCA data analyses (Markevitch et al. 1998; Irwin et
al. 1999).

The large negative declination of the bright radio source 
PKS~0316-444 associated with the cD galaxy, has adversely affected
the resolution of the VLA map (Takizawa et al. 2003). Nonetheless,
the map shows a strong nuclear component and two sharply bent radio 
tails with an overall extent approaching 100~kpc, similar to the 
radio galaxy in Abell~2029. The integrated spectral index of this source 
is $-0.8$ above 100~MHz.

\section{Observations and data analysis}

Abell~2029 was observed on April 12, 2003 with {\sl FUSE} through the LWRS 
aperture ($30\arcsec$$\times$$30\arcsec$) for 
a total time of 16~ksec (Program C0880201).
Abell~3112 was observed for a total time of 32~ksec on November 15 and 
November 16, 2002 (Program C0880101 and C0880102; 
for an overview of {\sl FUSE}, see Moos et al.~\cite{moos2000} and
Sahnow et al.~\cite{sahnow2000}).
The data of both targets were reprocessed with the version 2.2 of the 
{\tt CALFUSE} pipeline. 
The output of the pipeline is a total of 13 and 6~sub-exposures 
for Abell~3112 and Abell~2029, respectively. 
The sub-exposures have been aligned and co-added resulting
in a set of four independent spectra, one for each {\sl FUSE} channel 
(2 LiF spectra and 2 SiC spectra).
 
The resulting spectra are mainly composed of numerous terrestrial emission 
airglow lines, and O\,{\sc vi} solar contamination
(Sect.~\ref{galactic}). Despite a very low background, 
we do not detect any continuum, nor any signature of either C\,{\sc iii} or O\,{\sc vi} 
emissions from the clusters, or from their central cD galaxies. Thus, we derive
an upper limit of the order of a few
$10^{-15}$\,erg\,cm$^{-2}$\,s$^{-1}$\,\AA$^{-1}$ for the continuum
of each of the central galaxies. This non-detection of the
far-UV continuum is not surprising for these old ellipticals
at redshift $\sim 0.075$; a detection would require a large population
of several thousands of O~stars and a corresponding stellar formation
rate of several solar masses per year. The continuum flux of
the X-ray detected AGN in A3112 (Takizawa et al. 2003) 
is also far below the detection limit in the far-UV.

\begin{table*}
\caption{Characteristics of Abell~2029 and Abell~3112 and
parameters of the {\sl FUSE} observations.}
\begin{tabular}{lcccccccc}
\hline
\hline
\\
Cluster & $\alpha$ (2000) & $\delta$ (2000) & l      & b      & z & 
 $\lambda_{\rm OVI(1032)}$ & $\lambda_{\rm OVI(1038)}$ & Exp. Time \\
        &                 &                 & (deg.) & (deg.) &   & 
 (\AA)                     & (\AA)                     & (ks) \\
\hline
\\
Abell 2029 & 15h10m56s & 5\degr 44\arcmin 42\arcsec  &   6\fdg 4752 &  50\fdg 5463 & 0.0767 & 
 1111.1 & 1117.2 & 16\\
\\
Abell 3112 & 3h17m58s & -44\degr 14\arcmin 17\arcsec & 252\fdg 9367 & -56\fdg 0790 & 0.0746 &
 1108.9 & 1115.0 & 32 \\
\hline
\hline
\end{tabular}
\end{table*}

\begin{table*}
\label{Table 2}
\caption{Upper limits on the O\,{\sc vi} $\lambda$1032\AA\ emission line
and the corresponding mass flow rate}
\begin{tabular}{lccccc}
\hline
\hline
\\
Cluster & Size of aperture & N$_{\rm H I}$       & Attenuation$^{\rm a}$ & I\ {(1032\AA)} & $\dot{\rm M}$  \\
       & (kpc) & ($10^{21}$~cm$^{-2}$)& Factor               &(erg~~cm$^{-2}$~s$^{-1}$) &(M$_{\odot}$yr$^{-1}$) \\
\hline
\\
Abell 2029 & 67 & 0.31 & 2.0 & $<5\times 10^{-16}$ & $<27$ (3$\sigma$)\\
\\
Abell 3112 & 65 & 0.40 & 2.5 & $<4\times 10^{-16}$ & $<25$ (3$\sigma$)\\
\hline
\hline
\end{tabular}
\\
a: Extinction at $\sim$ 1100 \AA.\\
\end{table*}

\section{Results}

\begin{figure}
\includegraphics[width=\columnwidth]{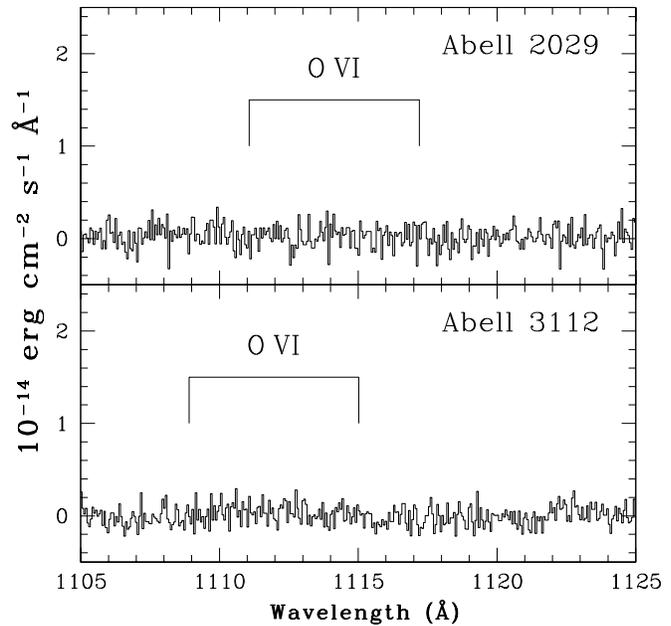}
\caption[]{The {\sl FUSE} spectra in the region of the O\,{\sc vi} doublet, 
with the expected positions of the two lines marked corresponding to the 
redshifts of Abell~2029 and
Abell~3112 ($z=0.0767$ and $z=0.0746$, respectively).  }
\label{plot_ovi_2029_3112}
\end{figure}

\begin{figure}
\includegraphics[width=\columnwidth]{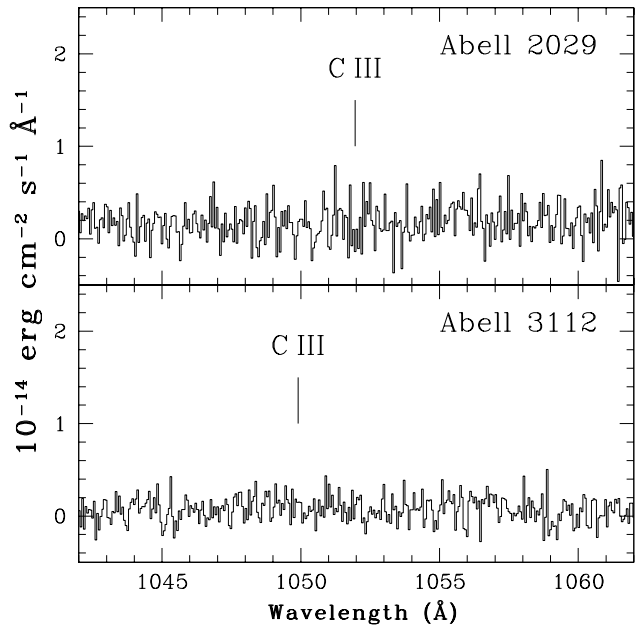}
\caption[]{Spectra of Abell~2029 and Abell~3112 in the region of the
C\,{\sc iii} 977\AA\ line with the line position expected at the
cluster redshifts. The C\,{\sc iii} line is 
not detected with an upper limit for the line intensity of 
$\sim 10^{-15}$\,erg\,cm$^{-2}$\,s$^{-1}$.  
}
\label{plot_ciii_2029_3112}
\end{figure}

We do not detect any emission from the O\,{\sc vi}
$\lambda\lambda$1032-1038\AA\ or C\,{\sc iii} $\lambda$977\AA\ lines from
the central $\sim$60\,kpc regions of Abell~2029 and Abell~3112
(Figs.~\ref{plot_ovi_2029_3112} and~\ref{plot_ciii_2029_3112}).
By fitting the spectra to the sum of a first-order polynomial for
the background and an emission line whose wavelength is defined by the
cluster redshift, we obtained the upper limits to the emission
intensities. The best-fit was always obtained for zero intensity of
the emission line. The emission line profile is calculated by the
convolution of a Gaussian with the instrumental response to a diffuse
emission filling the full LWRS aperture, which is a ``top-hat''
function with a width of $\sim$100\,km\,s$^{-1}$. Using various
values for the Gaussian FWHM between 0 and 200\,km\,s$^{-1}$, we
found that the estimated upper limit is not very sensitive to this width. The
3$\sigma$ upper limits are calculated by searching for the intensity
above which the $\chi^2$ of the fit is increased by at least 9.  We
thus obtained the upper limit for the O\,{\sc vi} line emission
at 1032~\AA : $I_{\rm Abell 2029} \la 5\times
10^{-16}$~erg~cm$^{-2}$~s$^{-1}$ and $I_{\rm Abell 3112} \la 4\times
10^{-16}$~erg~cm$^{-2}$~s$^{-1}$ (3$\sigma$).

The O\,{\sc vi} doublet being the main coolant at $\sim 3\times 10^5$\,K,  
the O\,{\sc vi} intensity is directly related to 
the cooling-flow rate through this temperature regime.
The upper limits given above can thus be translated into 
upper limits to the cooling-flow rates.
Using an intermediate case between isobaric and isochoric cooling,
the luminosity of the line at 1032\AA\ is given by 
$L \approx 0.9\cdot 10^{39} \dot{M_\odot}$~erg~s$^{-1}$, 
where $\dot{M_\odot}$ is the
mass flow rate in solar masses per year (Bregman et al. 2001; Oegerle et al.
2001).
Assuming that the cooling flow is fully covered by the 
{\sl FUSE} aperture, this gives the integrated line intensity within the 
{\sl FUSE} aperture of 
$I=2.2 \cdot 10^{-19}\dot{M_\odot}z^{-2}$~erg~cm$^{-2}$~s$^{-1}$, 
where z is the redshift of the corresponding cluster of galaxies. 

Finally, the upper limit for the O\,{\sc vi} line intensity must be corrected
for interstellar extinction.
This can be done using the published value of the H{\sc i} 
column densities (Table~2).
The observed upper limits for the
mass deposition rates are calculated after correcting for the attenuation factor 
$AF \equiv I_{\rm corrected}/I_{\rm observed}$.
Following  the galactic extinction law as given by 
Savage \& Mathis (1979), we assume that $AF\approx 10^{0.4 {A_\lambda}}$, 
where ${A_\lambda}\approx 11.55\, E(B-V)$ at $\lambda=1110$\,\AA\ 
and $E(B-V)\approx N($H\,{\sc i}$) / 4.8\times 10^{21}$cm$^{-2}$. 
This results in $(\log_{10} AF) \approx N($H\,{\sc i}$) / 10^{21}$cm$^{-2}$.
We thus obtain upper limits for the cooling flow rates of 
27 and 25\,M$_{\sun}$\,yr$^{-1}$ for Abell~2029 and Abell~3112, 
respectively (3$\sigma$ upper limits).

Absorption by Galactic H$_2$ at the wavelengths 
of the O\,{\sc vi} emission at the cluster redshifts 
could have been responsible for the non-detections in the two clusters.
However, absorption by Galactic H$_2$ appears unlikely. 
Although at $z\sim0.075$ O\,{\sc vi}\ $\lambda$1032\AA\ falls
within the H$_2$ $v$=0-0 band, the closest lines are the $R(0)$ 
and $R(1)$ lines ($\lambda_0$=1108.14\,\AA\ and 
$\lambda_0$=1108.64\,\AA) for Abell~2029 and the 
$P(1)$ and $R(2)$ lines 
($\lambda_0$=1110.07\,\AA\ and $\lambda_0$=1110.13\,\AA) for Abell~3112.
Very large column densities would be required to obtain a significant 
absorption of the O\,{\sc vi} emission at the cluster redshifts:
$N$(H$_2$,J=1)$\ga$2$\times 10^{21}$\,cm$^{-2}$ and 
$N$(H$_2$,J=2)$\ga 10^{21}$\,cm$^{-2}$ for Abell~2029, and 
$N$(H$_2$,J=0)$\ga 10^{21}$\,cm$^{-2}$ and 
$N$(H$_2$,J=1)$\ga$2$\times 10^{20}$\,cm$^{-2}$ for Abell~3112.
These column densities are clearly excluded toward these directions
with low H\,{\sc i} column densities (Table 2).

In summary, we obtained an upper limit of 
close to $\sim$25~M$_{\odot}$yr$^{-1}$ for the cooling-flow rates
in both clusters, within their central $\sim60$ kiloparsecs.

\section{Discussion}

The UV lines of the O\,{\sc vi}
$\lambda\lambda$1032-1038\AA\ doublet and/or of the C\,{\sc iii}
$\lambda$977\AA\ line offer a new tool and independent method to place
constraints on the cooling-flow rates in clusters of galaxies.
The upper limits for the cooling flow rates in Abell 2029 and Abell
3112, derived here from the UV resonance lines of O\,{\sc vi}
$\lambda\lambda$1032-1038\AA\ are 10 to 20 times lower than those
estimated earlier from ROSAT observations (Peres et al. 1998). The
discrepancy may actually be somewhat smaller in view of the possibility that
the cooling flow regions may have been covered only partially within the FUSE
aperture. Moreover, the O\,{\sc vi} lines may well be weakened due to
scattering by dust within the cluster cores (see Oegerle et
al. 2001). In any case, our upper limits are fully consistent with
those derived recently for the mass deposition rates of the keV gas,
using the Chandra and XMM-Newton observations of these clusters (Lewis
et al. 2002; Takizawa et al. 2003; see Sect. 2). Thus, the present results
indicate a general consistency for two separate temperature regimes of
the cooling ICM in these clusters. Begelman \& Fabian (1990) have
discussed the possibility that the $(1-3)\times 10^5$\,K gas could be rapidly
generated by turbulent mixing of the hot ($10^7$\,K) and warm ($10^4$\,K) 
phases. Thus, provided the intracluster
magnetic fields do not effectively suppress the mixing process, the
cooling may proceed too rapidly for a strong emission of the Fe\,{\sc xvii}
X-ray line. UV and X-ray observations therefore provide complementary
information on the radiative processes within the intracluster medium.

It may also be noted that the upper limits deduced
here for the O\,{\sc vi} $\lambda$1032\AA\ line intensities from the two clusters are
nearly half the intensity reported for Abell~2597 (Oegerle et
al. 1991). Since the redshift of Abell~2597 is
slightly larger than those of Abell~2029 and Abell~3112, and Galactic extinction is
similar, it is
clear that the O\,{\sc vi} emission from these two clusters 
and their corresponding cooling rates through $\sim 3\times 10^5$\,K 
are intrinsically lower as compared to Abell~2597.

\section{O\,{\sc vi} emission at rest-wavelength and the solar contamination}
\label{galactic}

\begin{figure}
\includegraphics[width=\columnwidth]{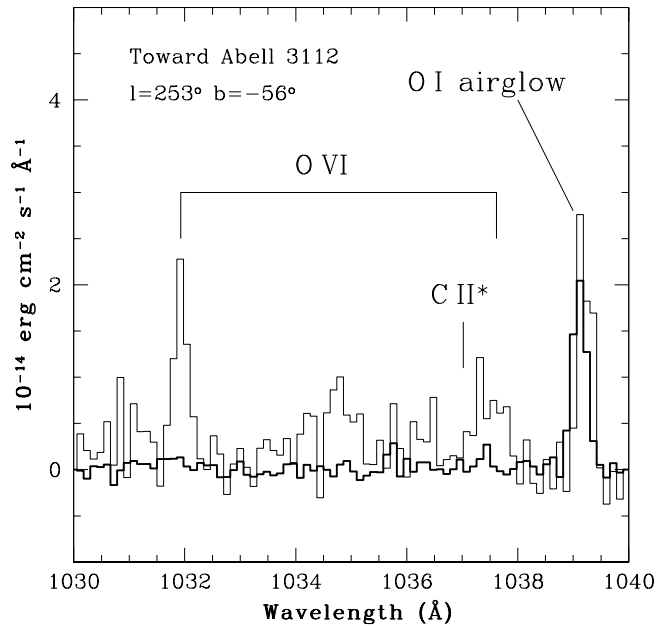}
\caption[]{Plot of the 1032-1038\AA\ O\,{\sc vi} doublet, showing solar contamination 
$(z = 0)$ during the observations of Abell~3112. The thin (thick) line shows the spectrum 
obtained by the co-addition of all SiC (LiF) data.
The difference between these two spectra 
clearly shows that the SiC channels are affected by 
O\,{\sc vi} Solar contamination.
}
\label{plot_ovi_sic_lif_3112}
\end{figure}

In the full spectrum of Abell~3112, obtained by combining the data from
the SiC and LiF channels, we serendipitously discovered 
an O\,{\sc vi} 1032--1038\AA\ emission doublet at zero radial velocity. 
Initially, we suspected this feature to originate from the interstellar
medium in our Galaxy, similar to the detections reported towards a few
other lines of sight (Shelton et al. 2001; Dixon et al. 2001b; Shelton
2002; Welsh et al. 2002; Otte et al. 2003).
This hypothesis also seemed consistent with the levels of
those detections being quite similar to the observed line intensities 
in our spectra:
$I(\lambda$1032\AA)$\sim$3460$\pm$740\,photons\,cm$^{-2}$\,s$^{-1}$\,sr$^{-1}$
and
$I(\lambda$1038\AA)$\sim$2220$\pm$990\,photons\,cm$^{-2}$\,s$^{-1}$\,sr$^{-1}$.
Moreover, the C\,{\sc iii} line at 1175.7\,\AA , which is often used
as a signature of solar contamination, was not present in our
spectra (see explanation below).  
However, a closer inspection of the spectra from the SiC and LiF
channels revealed that the O\,{\sc vi} doublet is only present in the SiC channels
(Fig.~\ref{plot_ovi_sic_lif_3112}).
This is a clear signature of 
a solar contamination, rather than emission from Galactic ISM, due to the fact
that the SiC mirrors are on the Sun-illuminated side of the spacecraft
(Shelton et al. 2001).
The absence of another known signature of solar contamination, the C\,{\sc iii} 
line at 1175.7\,\AA , can be readily understood by
recalling that the SiC channels do not extend to the wavelength of this line.
Clearly, this solar contamination must not be mistaken for Galactic emission.
Thus, it follows that the O\,{\sc vi} emission from the
Galactic interstellar medium remains undetected in the directions 
of the two clusters.

\section{Summary}

The non-detection of the O\,{\sc vi} $\lambda\lambda$1032-1038\AA\ and
C\,{\sc iii} $\lambda$977\AA\ lines from the clusters Abell~2029 and
Abell~3112, previously believed to have strong cooling-flow rates of
several hundred M$_\odot$yr$^{-1}$ adds to the growing independent
evidence that the magnitude of cooling flows in the cores of clusters
has been overestimated by at least an order of magnitude.

On a cautionary note we highlight the apparent detection of the 
O\,{\sc vi} doublet at about zero radial velocity in the spectrum taken
towards Abell~3112. The apparent emission
almost perfectly mimics emission
from the Galactic warm gas, similar to that already detected towards a 
few directions (e.g., Shelton et al. 2001). However, we argue that the
O\,{\sc vi} emission seen in the spectrum of Abell~3112, in fact,
arises from Solar contamination.

\begin{acknowledgements}
We warmly thank J.-M.~D\'esert for his help in data reduction,
G.~H\'ebrard for valuable comments on Solar contamination,
and A.~Vidal-Madjar for fruitful discussions.
We are grateful to the referee, Prof.\ J.~Bregman, for his constructive
remarks.
G-K thankfully acknowledges a travel grant from EGIDE, courtesy 
Prof.\ A.~Omont.
\end{acknowledgements}

\end{document}